\documentclass{ws-ijmpcs}
\newcommand{\be}{\begin{equation}}
\newcommand{\ee}{\end{equation}}
\newcommand{\bea}{\begin{eqnarray}}
\newcommand{\eea}{\end{eqnarray}}
\newcommand{\nn}{\nonumber}
\newcommand{\bnabla}{\mbox{\boldmath{$\nabla$}}}
\newcommand{\bGamma}{\mbox{\boldmath{$\Gamma$}}}
\newcommand{\bmalpha}{\mbox{\boldmath{$\alpha$}}}
\newcommand{\bPhi}{\mbox{\boldmath{$\Phi$}}}
\newcommand{\bmM}{\mbox{\boldmath{$\mathcal{M}$}}}
\newcommand{\bmN}{\mbox{\boldmath{$\mathcal{N}$}}}
\newcommand{\bmrho}{\mbox{\boldmath{$\rho$}}}
\newcommand{\bmtheta}{\mbox{\boldmath{$\theta$}}}
\newcommand{\bmone}{\mbox{\boldmath{$1$}}}
\begin{document}

\markboth{Milton, Abalo, Parashar, Pourtolami, Brevik, and Ellingsen}
{Repulsive Casimir Effects}

%
\catchline{}{}{}{}{}
%

\title{REPULSIVE CASIMIR EFFECTS}

\author{K. A. MILTON, E. K.  ABALO, PRACHI PARASHAR, NIMA POURTOLAMI}

\address{Homer L. Dodge Department of Physics and Astronomy,
University of Oklahoma, Norman, OK 73019 USA\\
milton@nhn.ou.edu, abalo@nhn.ou.edu, prachi@nhn.ou.edu, nimap@ou.edu}

\author{IVER BREVIK and S. \AA. ELLINGSEN}

\address{Department of Energy and Process Engineering,
Norwegian University of Science and Technology, N-7491
Trondheim, Norway\\
iver.h.brevik@ntnu.no, simen.a.ellingsen@ntnu.no}

\maketitle

\begin{history}
\received{Day Month Year}
\revised{Day Month Year}
\end{history}

\begin{abstract}

Like Casimir's original force between conducting plates in vacuum,
Casimir forces are usually attractive.
But repulsive Casimir forces can be achieved in special circumstances.
These might prove useful in nanotechnology.  We give examples of
when repulsive quantum vacuum forces can arise with conducting materials.

\keywords{Casimir forces, Casimir repulsion, Casimir-Polder interactions}
\end{abstract}

\ccode{PACS numbers: 42.50.Lc, 32.10.Dk, 12.20.-m, 03.50.De}

\section{Multiple Scattering Technique}
The multiple scattering approach starts from the well-known formula
for the electromagnetic quantum vacuum energy or Casimir energy 
($\tau$ is the ``infinite'' time that the
configuration exists)\cite{Schwinger-1975}
\be
E=\frac{i}{2\tau}\mbox{Tr}\,\ln \bGamma\to\frac{i}{2\tau} \mbox{Tr}\,\ln \bGamma\bGamma_0^{-1},
\ee
where $\bGamma$  is the Green's dyadic  satisfying
\be
\left(\frac1{\omega^2}\bnabla\times\frac1\mu\bnabla-\varepsilon\right)\bGamma=\bmone,
\ee
while $\bGamma_0$ satisfies the same equation with $\varepsilon=\mu=1$
everywhere.
We will choose the Green's dyadics satisfying outgoing-wave boundary
conditions (corresponding to the Feynman propagator), although other choices
can be made.

Consider material bodies characterized by a permittivity 
$\varepsilon(\mathbf{r})$ and a permeability $\mu(\mathbf{r})$,
so we have corresponding electric and magnetic potentials
\be
V_e(\mathbf{r})=\varepsilon(\mathbf{r})-1,\quad\mbox{and}\quad
V_m(\mathbf{r})=\mu(\mathbf{r})-1.
\ee
Then the trace-log appearing in the vacuum energy
is ($\bPhi_0=-\frac1\zeta\bnabla\times\bGamma_0$)
\bea
\mbox{Tr}\,\ln \bGamma\bGamma_0^{-1}&=&-\mbox{Tr}\,
\ln(\mathbf{1}-\bGamma_0 V_e)-\mbox{Tr}\,
\ln(\mathbf{1}-\bGamma_0 V_m)\nonumber\\
&&-\mbox{Tr}\,\ln(\mathbf{1}+\bPhi_0 \mathbf{T}_e\bPhi_0 \mathbf{T}_m),
\eea
in terms of the $\mathbf{T}$-matrix,
\be
\mathbf{T}_{e,m}=V_{e,m}(\mathbf{1}-\bGamma_0 V_{e,m})^{-1}.\ee
If we have {\em disjoint\/} electric bodies, the interaction term
separates out:
\bea
\mbox{Tr}\,\ln \left(\mathbf{1}-\bGamma_0(V_1+V_2)\right)&=&-\mbox{Tr}\,\ln
(\mathbf{1}-\bGamma_0\mathbf{T}_1)\nn\\
-\mbox{Tr}\,\ln
(\mathbf{1}-\bGamma_0\mathbf{T}_2)
&-&\mbox{Tr}\,\ln(\mathbf{1}-\bGamma_0\mathbf{T}_1
\bGamma_0\mathbf{T}_2),
\eea
so only the latter term contributes to the  interaction energy,
\be
E_{\rm int}=\frac{i}2\mbox{Tr}\ln(\mathbf{1}-\bGamma_0 \mathbf{T}_1
\bGamma_0 \mathbf{T}_2).
\ee
The same is true if one body is electric and the other magnetic,
\be
E_{\rm int}=-\frac{i}2\mbox{Tr}\ln(1+\bPhi_0 \mathbf{T}_1^e\bPhi_0
\mathbf{T}_2^m).\ee
Using this, it is straightforward
 to show that the Lifshitz energy per area between parallel
dielectric and diamagnetic  slabs, separated by a distance $a$, is
\be
\mathcal{E}_{\varepsilon\mu}=\frac1{16\pi^3}\int d\zeta\int d^2k\bigg[\ln\left(
1-r_1 r_2'e^{-2\kappa a}\right)
+\ln\left(1-r_1' r_2e^{-2\kappa a}\right)\bigg],\label{eem}\ee
where
\be
r_i=\frac{\kappa-\kappa_i}{\kappa+\kappa_i},\quad r_i'=
\frac{\kappa-\kappa'_i}{\kappa+\kappa'_i},\ee
with
\be\kappa^2=k^2+\zeta^2,\quad 
\kappa^2_1=k^2+\varepsilon\zeta^2, \quad \kappa_1'
=\kappa_1/\varepsilon,\quad \kappa^2_2=k^2+\mu\zeta^2,\quad\kappa_2'
=\kappa_2/\mu.
\ee
This means in the perfect reflecting limit,
$\varepsilon\to\infty$, $\mu\to\infty$,
\be
\mathcal{E}_{\rm Boyer}=+\frac78\frac{\pi^2}{720 a^3},
\ee
 we get Boyer's repulsive result.\cite{boyer74}

It is also well known, apparent from the purely electric version of
(\ref{eem}),
 in the Lifshitz-Dzyaloshinskii-Pitaevskii 
situation\cite{dlp}
of parallel dielectric media, with the intermediate medium having an 
intermediate value of the permittivity:
\be \varepsilon_1>\varepsilon_3>\varepsilon_2,
\ee
there is a Casimir repulsion between the upper and lower media.
This was demonstrated in the Munday-Capasso-Persegian experiment.\cite{munday}

\section{Casimir Effect on Spheres and Cylinders}
Earlier Boyer had shown\cite{boyer} that the Casimir self-energy of a spherical
shell was positive, that is, repulsive.  Such calculations have been
generalized, as displayed in Table \ref{tab1}.  Note that all these
energies for spheres are positive (repulsive). 
Very recently, energies for cylinders with Dirichlet boundaries
having triangular cross sections were computed, which also
display positive Casimir self-energies,\cite{abalo} as we discuss in
Sec.~\ref{sec4} below.
\begin{table}[ph]
\tbl{Casimir energy ($E$) for a sphere and Casimir energy per unit
length ($\mathcal{E}$) for an infinite cylinder, both of radius $a$.
Here the different boundary
conditions are perfectly conducting for electromagnetic fields (EM),
Dirichlet for scalar fields (D), dilute dielectric for electromagnetic
fields [coefficient of $(\varepsilon-1)^2$], dilute dielectric for
electromagnetic fields
with media having the same speed of light (coefficient of $\xi^2
=[(\varepsilon-1)/(\varepsilon+1)]^2$),
and weak coupling
for scalar field with $\delta$-function boundary
(coefficient of $\lambda^2/a^2$).  The references given are, to the authors'
knowledge, the
first paper in which the results in the various cases were found.}
{\begin{tabular}{p{2cm}p{4cm}p{3cm}}
\toprule
Type&$E_{\rm Sphere}a$&$\mathcal{E}_{\rm Cylinder}a^2$\\
\colrule
EM&$+0.04618$\cite{boyer} &$-0.01356$\cite{DeRaad:1981hb}\\
D&$+0.002817$\cite{Bender:1994zr}&$+0.0006148$\cite{gosrom}\\
$(\varepsilon-1)^2$&$+0.004767=\frac{23}{1536\pi}$\cite{Brevik:1998zs}
&$0$\cite{Cavero-Pelaez:2004xp}\\
$\xi^2$&$+0.04974=\frac5{32\pi}$\cite{Klich:1999df}&$0$\cite{nesterenko}\\
$\lambda^2/a^2$&$+0.009947=\frac1{32\pi}$\cite{Milton:2002vm}
&$0$\cite{CaveroPelaez:2006rt}\\
\botrule
\end{tabular}
\label{tab1}}

\end{table}


\section{Dimensional Dependence}

\begin{figure}
\centerline{
\psfig{figure=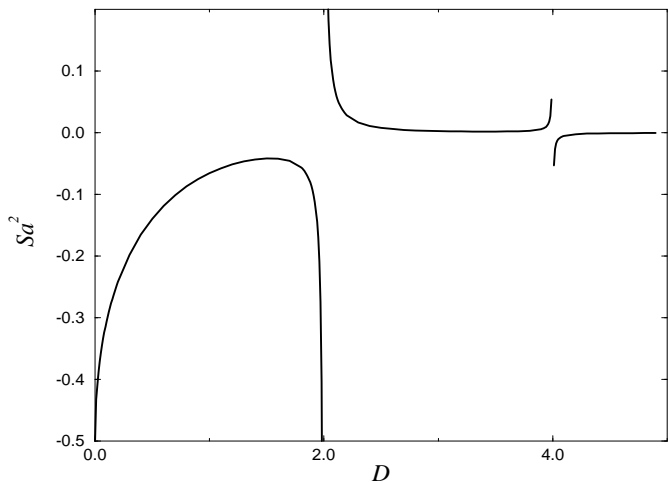,%
height=4in,width=3in,angle=270}}
\caption{\label{figd}
Scalar Casimir stress ${\cal S}$ for $0<D<5$ on a spherical shell.
imposing Dirichlet boundary conditions. Here $D$ is the number of spatial dimensions.}
\end{figure}

Bender and Milton\cite{Bender:1994zr}
considered the Casimir effect due to fluctuations in a scalar
field interior and exterior to a Dirichlet hypersphere, in $D$ spatial
dimensions, and found
that poles occur in even spatial dimensions, as shown in Fig.~\ref{figd},
which shows that for a scalar field, subject to Dirichlet boundary
conditions on the hyperspherical surface, repulsion occurs for
$2<D<4$.

\section{Triangular Cylinders}\label{sec4}
For an equilateral triangle of height $h$,
the scalar eigenmodes corresponding to Dirichlet
boundary conditions are known explicitly\cite{embook,radbook}
\be
\gamma_l^2=\frac23\left(\frac{\pi}{h}\right)^2(l_1^2+l_2^2+l_3^2),
\quad
l_1+l_2+l_3=0,\quad l_i\ne0.
\ee
Although this has been appreciated for most of a century, only
last year was the corresponding calculation of the Casimir energy
for a cylinder with such a triangular cross section (and also
those obtained by bisecting an equilateral triangle and a square)
carried out.\cite{abalo}

In $d$ longitudinal dimensions, the Casimir energy is
\be
\mathcal{E}=-\frac{\Gamma(-1/2-d/2)}{2^{2+d}\pi^{(d+1)/2}}
\sum_l(\gamma_l^2)^{(d+1)/2},
\ee
which can be analytically continued and summed by means of the
exceedingly rapidly convergent Chowla-Selberg
formula,\cite{elizalde}
\be
\mathcal{E}=+\frac{0.0177891}{h^2}.
\ee

 We can also evaluate the eigenvalue sum by use of the
Poisson sum formula,
\be
\sum_{l=-\infty}^\infty f(l)=2\pi\sum_{k=-\infty}^\infty \tilde f(k),
\ee
in terms of the Fourier transform
\be
\tilde f(k)=\int_{-\infty}^\infty \frac{d\alpha}{2\pi} e^{2\pi i k\alpha}
f(\alpha).
\ee
We use the Poisson sum formula together with
(temporal)  point-splitting regularization,
starting from
\bea
E=\frac1{2i}\int(d\mathbf{r})\int\frac{d\omega}{2\pi}2\omega^2\mathcal{G}
(\mathbf{r,r})e^{-i\omega\tau},
\eea
with $\tau\to0$, which for a cylindrical waveguide gives
for the energy per unit length
\bea
\mathcal{E}
&=&\frac12\int_{-\infty}^\infty \frac{d\zeta}{2\pi}2(-\zeta^2)\int
\frac{dk}{2\pi}\sum_{m,n}\frac1{\zeta^2+k^2+\gamma_{mn}^2}e^{i\zeta\tau}\nn\\
&=&\frac{1}{2}\, \lim_{\tau\to0} \left(- \frac{d}{d\tau} \right)
\int_{-\infty}^{\infty} \frac{dk}{2\pi}\,
\sum_{m,n} e^{-\tau \sqrt{k^2+ \gamma_{m n}^{2}}}.\label{e-ptsplt}
\eea

A virtue of the point-splitting method is that
we can isolate the divergences in the energy:
\be
\mathcal{\widehat{E}}_{\rm Eq}^{(D)}
=\lim_{\tau\to0}\left(\frac{3A}{2\pi^2\tau^4}-\frac{P}{8\pi\tau^3}
+\frac1{6\pi\tau^2}
\right).\label{div-e}
\ee
We note that the ``volume'' and ``surface'' divergent terms,
which are respectively proportional to
the area of the triangle $A=h^2/\sqrt{3}$  and the
perimeter $P=2\sqrt{3}h$, are as expected,
and are presumably not of physical relevance.
The last term, a constant in $h$, certainly does
not contribute to the self-stress on the cylinder. Only this term reflects
the corner divergences.  For a general polygon, with interior angles
$\alpha_i$, the last term is
\be
\frac{1}{48\pi}\sum_i \left(\frac\pi{\alpha_i}-\frac{\alpha_i}\pi\right)
\frac1{\tau^2}.
\ee  These coefficients are proportional to the heat kernel 
coefficients---in particular there is no $a_2$ 
heat kernel coefficient, because the surfaces are flat,
 which means that the 
Casimir energy can be identified unambiguously.

Remarkably, for the integrable polygonal figures we are considering,
the Casimir energy can be given in closed form,  
in terms of the polygamma function.  Thus
\bea
\mathcal E_{\rm Eq}^{(D)}&=&-\frac1{96 h^2}\left[\frac{\sqrt{3}}9\left[
\psi'(1/3)-\psi'(2/3)\right]-\frac8\pi \zeta(3)\right]
=\frac{0.0177891}{h^2}.
\eea
It is {\it a priori\/} remarkable that such an explicit form can be achieved
for a strong-coupling problem.

The same methods can be used to evaluate the Casimir energy
for a square wave\-guide (side $a$), a well-studied
system,\cite{lukosz,aw} although the closed form was
previously unknown,
\bea
\mathcal{E}_{\rm Sq}^{(D)}&=&-\frac{1}{32\pi^2a^2}\bigg[2\zeta(4)-\pi\zeta(3)
+8\pi^2\sum_{l=1}^\infty l^{3/2}\sigma_3(l)K_{3/2}(2\pi l)
\bigg]\nn\\
&=&-\frac1{32\pi^2a^2}\bigg[4\zeta(4)-2\pi\zeta(3)
+4\sum_{k,l=1}^\infty\frac{1}{(k^2+l^2)^2}\bigg]\nn\\
 &=&
\frac1{16\pi a^2}\left[\zeta(3)-\frac\pi3 G\right]=
\frac{0.00483155}{a^2}.
\eea

 By bifurcating the square, we can obtain the isosceles right
triangle, and by bifurcating the equilateral triangle we can
get the $30^\circ$-$60^\circ$-$90^\circ$ triangle:
\bea
\mathcal{E}_{\rm Iso}^{(D)}&=&\frac12\mathcal{E}_{\rm Sq}^{D)}
+\frac{\zeta(3)}{16\pi a^2}
=\frac{0.0263299}{a^2},\\
\mathcal{E}_{\rm 369}^{(D)}&=&\frac12\mathcal{E}_{\rm Eq}^{(D)}
+\frac{\zeta(3)}{8\pi h^2}
=\frac{0.0567229}{h^2},\eea
to be compared to the result for a circle\cite{gosrom}
\be
\mathcal{E}_{\rm Circ}^{(D)}=\frac{0.0006148}{a^2}.
\ee
For the latter, external contributions must be included, to cancel
the curvature divergences. 

We can also get results for Neumann boundary conditions
(H or TE modes)
\bea
\mathcal{E}^{(N)}_{\rm Sq}&=&\mathcal{E}^{(D)}_{\rm Sq}-\frac{\zeta(3)}{8\pi a^2}
=-\frac{0.0429968}{a^2},\\
\mathcal{E}^{(N)}_{\rm Eq}&=&\mathcal{E}^{(D)}_{\rm Eq}-\frac{\zeta(3)}{6\pi h^2}
=-\frac{0.045982}{h^2},\\
\mathcal{E}^{(N)}_{\rm Iso}&=&\frac12\mathcal{E}^{(N)}_{\rm Sq}-\frac{\zeta(3)}{16\pi a^2}
=-\frac{0.0454125}{a^2},\\
\mathcal{E}^{(N)}_{\rm 369}&=&\frac12\mathcal{E}^{(N)}_{\rm Eq}-\frac{\zeta(3)}{8\pi h^2}
=-\frac{0.0708193}{h^2}.\eea

\begin{figure}
\centering
\includegraphics[scale=.5]{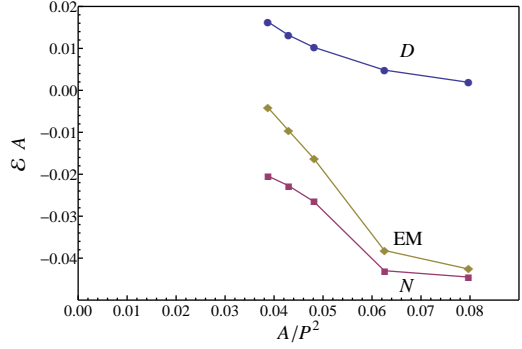}
\caption{\label{figsys} Dependence of Casimir energies of triangular 
waveguides.  The top curve is for Dirichlet boundary conditions,
the bottom for Neumann boundary conditions, and the intermediate
curve represents surfaces which are perfect electromagnetic conductors.}
\end{figure}

\begin{figure}
\begin{center}
\epsfig{file=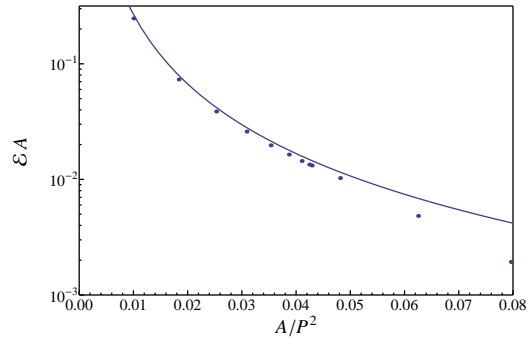,height=5cm}
\end{center}
\caption{\label{fig-cyl-pl} 
Scaled Dirichlet Casimir energies for
triangular waveguides 
both for numerical and analytically solvable cases.  The two rightmost points 
correspond to a square and a circular cross section.
 The solid line is the PFA approximation.}
\end{figure}

Graph \ref{figsys} shows the systematic dependence of $\mathcal{E}^{(D)}$,
$\mathcal{E}^{(N)}$ and $\mathcal{E}^{({\rm EM})}$
expressed in the dimensionless form $\mathcal{E}A$ in terms of
the geometrical quantity $(A/P^2)$, where $A$ is the cross-sectional area,
and $P$ is the cross-sectional perimeter of the waveguide.
The limited analytic results have been supplemented by
a numerical method to extract eigenvalues
for right triangles with arbitrary acute angles.  Those results,
shown in Fig.~\ref{fig-cyl-pl} for the Dirichlet case,
lie on our universal curve, and agree with the proximity force
approximation PFA (solid line) for small
acute angles:
\be
\mathcal{E}^{(D)}_{\rm PFA}A=-\frac{\pi^2}{1440}\int_0^a \frac{dr}{(r\theta)^3}\frac12 a^2\theta
\to \frac{\pi^2}{368640}\left(\frac{P^2}A\right)^2.
\ee

\section{Classical Repulsion}
Both classical and quantum repulsion were described last
year by Levin et al.\cite{levin}, and we give some
additional examples here.  (More details of our considerations
appear in Ref.~\refcite{rep1}.)

\subsection{Classical dipole interaction}
\label{sec3}
\begin{figure}
 \begin{center}
\includegraphics[scale=1]{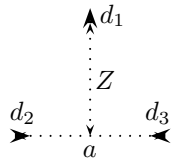}
\caption{\label{fig3} Configuration of three dipoles, two of which are
antiparallel, and one perpendicular to the other two.}
\end{center}
\end{figure}

It is possible to achieve a repulsive force between a configuration of
fixed dipoles.  Consider the situation illustrated in Fig.~\ref{fig3}.
Here we have two dipoles, of strength $d_2$ and $d_3$ lying along the $x$
axis, separated by a distance $a$.  A third dipole of strength $d_1$ lies
along the $z$ axis.  If the two parallel dipoles are oppositely directed
and of equal strength,
\be
\mathbf{d}_2=-\mathbf{d}_3=d_2 \mathbf{\hat x},
\ee
 equally distant from the $z$ axis,
and the dipole on the $z$ axis is directed along that axis,
\be
\mathbf{d}_1=d_1\mathbf{\hat z},
\ee
the force on that dipole is along the $z$ axis:
\be
F_z=3ad_1d_2\frac{a^2/4-4 Z^2}{(Z^2+a^2/4)^{7/2}},
\ee
which changes sign at $Z=a/4$; that is, close to $Z=0$ the force
on dipole 1 is repulsive. 

\subsection{Interaction of atom with aperture}
\begin{figure}
 \begin{center}
\includegraphics[scale=1]{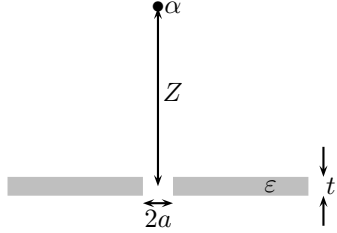}
\caption{\label{figap} Three-dimensional geometry of a dipole or polarizable
atom a distance $Z$ above a dielectric or conducting plate with a circular
aperture of radius $a$.  In this paper we will consider a conductor with
thickness $t=0$.}
\end{center}
\end{figure}

We first consider a dipole above an aperture in a perfectly
conducting line in two dimensions, as shown in Fig.~\ref{figap}.
The Green's function which vanishes on the entire line $z=0$ is
\be
G(\mathbf{r,r'})=-\ln[(x-x')^2+(z-z')^2]+\ln[(x-x')^2+(z+z')^2],
\ee
with the boundary condition:
\be G(x,0;x',z')=0.
\ee
Then the electrostatic potential at any point above the $z=0$ line is
\be
\phi(\mathbf{r})=\int_{z>0}(d\mathbf{r'})G(\mathbf{r,r'})\rho(\mathbf{r'})
+\frac1{4\pi}\int_{\rm ap}dS'\frac\partial{\partial z'}
G(\mathbf{r,r'}) \phi(\mathbf{r'}),
\ee
where
\be
\rho(\mathbf{r})=-\mathbf{d}\cdot\bnabla\delta(\mathbf{r-R}),\quad 
\mathbf{R}=(0,Z).
\ee
The surface integral extends only over the aperture because the potential
vanishes on the conducting line.  If we choose $\mathbf{d}$ to point along
the $z$ axis we easily find ($2a = $ width of aperture)
\bea
\phi(x,z>0)&=&2d\left[\frac{z-Z}{x^2+(z-Z)^2}+ \frac{z+Z}{x^2+(z+Z)^2}\right]
\nn\\
&&\mbox{}+
\frac1\pi\int_{-a}^{a} dx'\frac{z}{(x-x')^2+z^2}\phi(x',0).
\label{potgf}
\eea

Now the free Green's function in two dimensions is
\be
G_0(\mathbf{r,r'})=4\pi\int\frac{(d\mathbf{k})}{(2\pi)^2}\frac{e^{ik_x(x-x')}
e^{ik_z(z-z')}}{k_x^2+k_z^2}
=\int_{-\infty}^\infty dk_x\frac1{|k_x|}e^{ik_x(x-x')}e^{-|k_x||z-z'|}.
\ee
Then the surface integral above is
\be
\int_{-\infty}^\infty \frac{dk_x}{2\pi}e^{ik_x x}e^{-|k_x|z}\tilde\phi(k_x),
\ee
in terms of the Fourier transform of the field
\be
\tilde \phi(k_x)=\int_{-\infty}^\infty dx' e^{-ik_x x'}\phi(x',0)
=2\int_0^{a}dx'\cos k_xx'\phi(x',0),
\ee
since $\phi(x,0)$ must be an even function for the geometry considered.
Thus we conclude
\be
\phi(x,z>0)=2d\left[\frac{z-Z}{x^2+(z-Z)^2}+\frac{z+Z}{x^2+(z+Z)^2}\right]
+\frac1\pi\int_0^\infty dk\cos kx \,e^{-kz}\tilde\phi(k).
\ee
The electric field in the aperture is
\be
E_z(x,z=0+)=-\frac\partial{\partial z}\phi(x,z)\bigg|_{z=0+}
=-4d\frac{x^2-Z^2}{(x^2+Z^2)^2}+\frac1\pi\int_0^\infty dk\,k\cos kx\,
\tilde\phi(k).
\ee
On the other side of the aperture, there is no charge density, so for
$z<0$ the potential is
\be
\phi(x,z<0)=\frac1\pi\int_0^\infty dk\cos kx \,e^{kz}\tilde\phi(k),
\ee
so the $z$-component of the electric field in the aperture is
\be
E_z(x,z=0-)=-\frac\partial{\partial z}\phi(x,z)\bigg|_{z=0-}
=-\frac1\pi\int_0^\infty dk\,k\cos kx\,\tilde\phi(k).
\ee
Because we require that the electric field be continuous in the aperture,
and the potential vanish on the conductor, we obtain the two coupled
integral equations for this problem,
\bea
4d\frac{x^2-Z^2}{(x^2+Z^2)^2}&=&\frac2\pi\int_0^\infty dk\,k\cos kx\,
\tilde\phi(k),\quad 0<|x|<a,\\
0&=&\int_0^\infty dk\cos kx\,\tilde\phi(k),\quad |x|>a.
\eea
In fact, these equations have a simple solution\cite{Khanzhov}
\be
\tilde\phi(k)=-\frac{2Zd\pi}a\int_0^1dx\,x\frac{J_0(kax)}
{(x^2+Z^2/a^2)^{3/2}}.
\ee
From this, we can work out the energy of the system from
\be
U=-\frac12dE_z(0,Z)=\frac12d\frac{\partial\phi}{\partial z}\bigg|_{z=Z,x=0},
\label{energy}
\ee
where the factor of 1/2 comes from the fact that this must be the energy
required to assemble the system.  We must further
drop the self-energy of the dipole due to its own field.  
We are then left with
\bea
U_{\rm int}&=&-\frac{d^2}{4Z^2}-\frac{d}{2\pi}\int_0^\infty dk\,k\, 
e^{-kZ}\tilde
\phi(k)\nn\\
&=&-\frac{d^2}{4Z^2}+\frac{Z^2d^2}{a^4}\int_0^1\frac12 dx^2
\frac1{(x^2+Z^2/a^2)^3}\nn\\
&=&-\frac14\frac{Z^2d^2}{(a^2+Z^2)^2},
\eea
twice  that of Levin et al.\cite{levin}
Since this vanishes at $Z=0$ and $Z=\infty$, the force must change from
attractive to repulsive, which happens at $Z=a$.

\subsection{Circular aperture interacting with dipole}\label{sec4b}
It is quite straightforward to repeat the above calculation in three
dimensions.  Again we are considering a dipole, polarized on the symmetry
axis, a distance $Z$ above a circular aperture of radius
$a$  in a conducting plate, illustrated in Fig.~\ref{figap}.

The free three-dimensional
 Green's function in cylindrical coordinates has the representation
\be
\frac1{\sqrt{\rho^2+z^2}}=\int_0^\infty dk\,J_0(k\rho)e^{-k|z|},\label{freegcc}
\ee
and so if we follow the above procedure we find for the potential above
the plate
\bea
\phi(\mathbf{r_\perp},z>0)&=&d\left[\frac{z-Z}{[r_\perp^2+(z-Z)^2]^{3/2}}+
\frac{z+Z}{[r_\perp^2+(z+Z)^2]^{3/2}}\right]\nn\\
&&\quad\mbox{}+\int_0^\infty dk\,k\, e^{-kz}J_0(kr_\perp)\Phi(k),
\eea
where the Bessel transform of the potential in the aperture is
\be
\Phi(k)=\int_0^\infty d\rho\,\rho\, J_0(k\rho)\phi(\rho,0).
\ee

Thus the integral equations resulting from the continuity of the $z$-component
of the electric field in the aperture and the vanishing of the potential on
the conductor are
\bea
d\frac{r_\perp^2-2Z^2}{[r_\perp^2+Z^2]^{5/2}}&=&\int_0^\infty dk \,k^2 J_0(k
r_\perp)\Phi(k),\quad r_\perp<a,\\
0&=&\int_0^\infty dk\,k J_0(kr_\perp)\Phi(k),\quad r_\perp>a.
\eea
The solution to these equations is given in Titchmarsh's book,\cite{titchmarsh}
and after a bit of manipulation we obtain
\be
\Phi(k)=-\left(\frac{2ka}\pi\right)^{1/2}\frac{2dZ}{ka^2}\int_0^1 dx\,x^{3/2}
\frac{J_{1/2}(xka)}{(x^2+Z^2/a^2)^2}.
\ee
Then the energy  may be easily evaluated using
\be
\int_0^\infty dk\,k^{3/2}e^{-kZ}J_{1/2}(kax)=2\sqrt{\frac{2xa}\pi}
\frac{Z}{(x^2a^2+Z^2)^2}.
\ee
The energy can again be expressed in closed form:
\be
U=-\frac{d^2}{8Z^3}+\frac{d^2}{4\pi Z^3}\bigg[\arctan\frac{a}Z
+\frac{Z}a\frac{1+8/3(Z/a)^2-(Z/a)^4}{(1+Z^2/a^2)^3}\bigg].
\ee
This is always negative, but vanishes at infinity and at zero.
Numerically, we find that the force changes
sign at $Z=0.742358a$.
The reason why the energy vanishes when the dipole is centered in the
aperture is clear: Then the electric field lines are perpendicular to the
conducting sheet on the surface, and the sheet could be removed without
changing the field configuration.

\section{Casimir-Polder Energy}

Our goal now is to find analytically  the quantum (Casimir) analog of
this classical repulsion.
This was given a numerical study in Ref.~\refcite{levin}.
Here we want to offer an analytic counterpart.  We will in this
section be considering the Casimir-Polder (CP) interaction of an
atom with a conducting body, which is in general given by
\be
U_{\rm CP}=-\int_{-\infty}^\infty d\zeta \,\mbox{tr}\,
\bmalpha\cdot\bGamma(\mathbf{r,r};\zeta),
\ee
where $\bmalpha$ is the polarizability dyadic of the atom, located
at position $\bf r$, 
and $\bGamma$ is the Green's dyadic for the electromagnetic field
at imaginary frequency $\zeta$ corresponding to the conducting
body.

\subsection{Casimir-Polder force due to a conducting wedge}
\begin{figure}
 \begin{centering}
\includegraphics[scale=1]{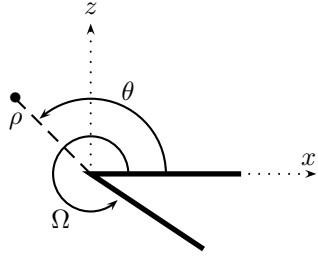}
\caption{\label{figwedge} Polarizable atom, located at polar coordinates
$\rho$, $\theta$, within a conducting wedge with dihedral angle $\Omega=\pi/p$.}
 \end{centering}
\end{figure}

Consider a polarizable atom located outside a conducting wedge, as shown
in Fig.~\ref{figwedge}.
The interaction between a polarizable atom and a perfectly conducting
half-plane is a special case of the vacuum interaction between such an atom
and a conducting wedge.  For an isotropic atom, the wedge was
considered by Brevik, Lygren, and Marachevsky.\cite{blm} 
In terms of the exterior dihedral angle of the wedge $\Omega$,
with $p=\pi/\Omega$, the electromagnetic Green's
dyadic has the form (here the translational direction is denoted by $y$,
and one plane of the wedge lies in the $z=0$ plane, the other intersecting
the $xz$ plane on the line $\theta=\Omega$---see Fig.~\ref{figwedge})
\bea
\bGamma(\mathbf{r,r'})&=&2p\sum_{m=0}^\infty{}'\int\frac{dk}{2\pi}
\bigg[-\bmM\bmM^{\prime*}(\nabla_\perp^2-k^2)\nn\\
&&\quad\times\frac1{\omega^2}
F_{mp}(\rho,\rho')\frac{\cos mp\theta \cos mp\theta'}\pi e^{ik(y-y')}  \nn\\
&&\mbox{}+\bmN\bmN^{\prime*}\frac1{\omega}
G_{mp}(\rho,\rho')\frac{\sin mp\theta \sin mp\theta'}\pi e^{ik(y-y')}\bigg].
\eea
The first term here refers to TE (H) modes, the second to TM (E) modes.
The prime on the summation sign means that the $m=0$ term is counted with
half weight.  In the polar coordinates in the $xz$ plane, $\rho$ and $\theta$,
the H and E mode operators are
\be
\bmM=\hat \bmrho\frac\partial{\rho\partial\theta}
-\hat\bmtheta\frac\partial{\partial \rho},\quad
\bmN= ik\left(\hat\bmrho\frac\partial{\partial\rho}
+\hat\bmtheta\frac\partial{\rho\partial \theta}\right)
-\mathbf{\hat y}\nabla_\perp^2.
\ee
where the transverse Laplacian is
\bea
\nabla_\perp^2=
\frac1\rho\frac\partial{\partial \rho}\rho\frac\partial{\partial\rho}
+\frac1{\rho^2}
\frac{\partial^2}{\partial \theta^2}.
\eea
In this situation, the boundaries are entirely in planes of constant $\theta$,
so the radial Green's functions are equal to the free Green's function
\be
\frac1{\omega^2}F_{mp}(\rho,\rho')=\frac1\omega G_{mp}(\rho,\rho')
=-\frac{i\pi}{2\lambda^2}
J_{mp}(\lambda \rho_<)H^{(1)}_{mp}(\lambda \rho_>),
\ee
with $\lambda^2=\omega^2-k^2$.
We will immediately make the Euclidean rotation, $\omega\to i\zeta$, where
$\lambda\to i\kappa$, $\kappa^2=\zeta^2+k^2$, so the free Green's functions
become  $-\kappa^{-2}I_{mp}(\kappa\rho_<)K_{mp}(\kappa \rho_>)$.

\subsubsection{Completely anisotropic atom}
We start by considering the most favorable case for 
Casimir-Polder repulsion, where only $\alpha_{zz}\ne0$.
In the static limit, where the frequency dependence of the
polarizability is neglected, then the only component of the Green's dyadic that
contributes comes in as 
\bea
\int\frac{d\zeta}{2\pi}\Gamma_{zz}
&=&\frac{2p}{4\pi^3}\int dk\,d\zeta
\bigg\{\left[\zeta^2\sin^2\theta\sin^2mp\theta-k^2\cos^2\theta\cos^2mp\theta
\right]\nn\\
&&\qquad\times\frac{m^2p^2}{\kappa^2\rho_<\rho_>}I_{mp}(\kappa\rho_<)
K_{mp}(\kappa\rho_>)\nn\\
&-&\left[k^2\sin^2\theta\sin^2 mp\theta-\zeta^2\cos^2\theta
\cos^2mp\theta\right] I'_{mp}(\kappa\rho_<)K'_{mp}(\kappa \rho_>)\bigg\}.
\eea
Here we note that the off diagonal $\rho$-$\theta$ terms in $\bGamma$ cancel.
We have regulated the result by point-splitting in the radial coordinate.
At the end of the calculation, the limit $\rho_<\to\rho_>=\rho$ is to be
taken.
Now the integral over the Bessel functions is given by
\be
\int_0^\infty d\kappa\,\kappa\, I_\nu(\kappa\rho_<)K_\nu(\kappa\rho_>)
=\frac{\xi^\nu}{\rho_>^2(1-\xi^2)},
\ee
where $\xi=\rho_</\rho_>$. After that the $m$ sum is easily carried out
by summing a geometrical series.  Care must also be taken with the $m=0$ term
in the cosine series.  The result of a straightforward calculation leads to
\be
\label{vacuum}
\int\frac{d\zeta}{2\pi}\Gamma_{zz}=-\frac{\cos 2\theta}{\pi^2\rho^4}
\frac1{(\xi-1)^4}+\mbox{finite}.
\ee
The divergent term is that of the vacuum without the wedge, so
we must subtract this term
off, leaving for the static Casimir-Polder energy
\be
U^{zz}_{\rm CP}=-\frac{\alpha_{zz}(0)}{8\pi}\frac1{\rho^4\sin^4p\theta}
\bigg[p^4-\frac{2}3p^2(p^2-1)\sin^2p\theta
+\frac{(p^2-1)(p^2+11)}{45}\sin^4p\theta
\cos2\theta\bigg].\label{ucpzz}
\ee
This result may also be easily derived from the closed form given by
Lukosz.\cite{lukosz}

A small check of this result is that as $\theta\to 0$ (or $\theta\to\Omega$)
we recover the expected Casimir-Polder result for an atom above an infinite
plane:
\be
U_{\rm CP}^{zz}\to -\frac{\alpha_{zz}(0)}{8\pi Z^4},
\ee
in terms of the distance of the atom above the plane, $Z=\rho\theta$.
This limit is also obtained when $p\to1$, for when $\Omega=\pi$ we are
describing a perfectly conducting infinite plane.

A very similar calculation gives the result for
an isotropic atom, $\bmalpha=\alpha{\bf1}$, which was first given by Brevik,
Lygren, and Marachevsky:\cite{blm}
\be
U_{\rm CP}=-\frac{3\alpha(0)}{8\pi\rho^4\sin^4p\theta}\bigg[p^4-\frac23p^2(p^2-1)
\sin^2p\theta-\frac13\frac1{45}(p^2-1)(p^2+11)\sin^4 p\theta\bigg].
\ee
Note that this is not three times $U_{\rm CP}^{zz}$ in above,
 because the $\cos 2\theta$ factor in the last term in the latter is replaced
by $-1/3$ here.  
\begin{figure}
 \begin{centering}
\includegraphics[scale=1]{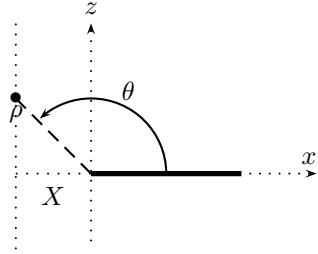}
\caption{\label{fighp}  Polarizable atom, above a half conducting plane,
free to move on a line perpendicular to the plane but a distance $X$ to
the left of the plane.}
 \end{centering}
\end{figure}
\subsubsection{Repulsion by half-plane}
Let us consider the special case $p=1/2$, that is $\Omega=2\pi$, the case of
a semi-infinite conducting plane, illustrated in Fig.~\ref{fighp}. 
This was the situation considered,
for anisotropic atoms, in recent papers  by Eberlein and Zietal.\cite{zietal1,zietal2}
 Consider a particle free to move along a
line parallel to the $z$ axis, a distance $X$ to the left of the
semi-infinite plane.  

The half-plane $x<0$
constitutes an aperture of infinite width. With $X$ fixed, we
can describe the trajectory by $u=X/\rho=-\cos\theta$,
which variable ranges from zero to one.  The polar angle is given by
\be
\sin^2\frac\theta2=\frac{1+u}2.
\ee
 The energy for an isotropic atom is given by
\be
U_{\rm CP}=-\frac{\alpha(0)}{32\pi}\frac1{X^4}V(u),
\ee
where
\be
V(u)=3u^4\left[\frac1{(1+u)^2}+\frac1{u+1}+\frac14\right].\label{vee}
\ee

\begin{figure}
 \begin{center}
\includegraphics[scale=1]{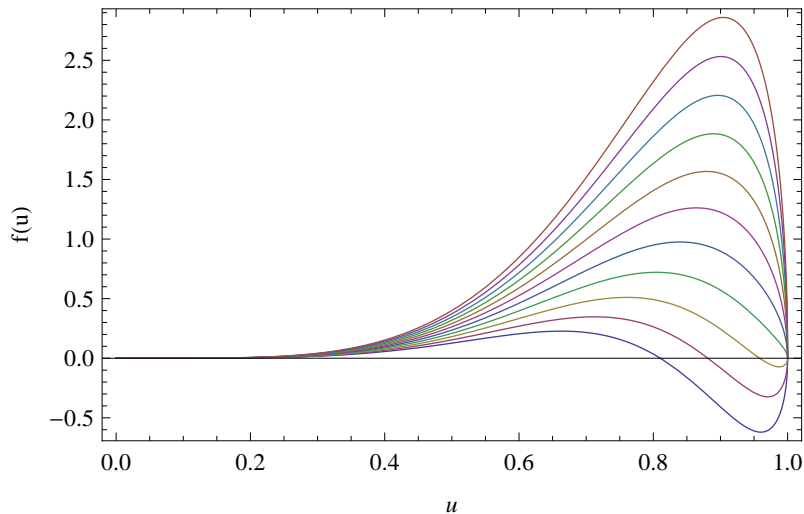}
\caption{\label{figsemi1} The $z$ component of the force between an
anisotropic atom and a semi-infinite conducting plane.
$F_z=-\alpha_{zz}/(32\pi X^5) f(u)$ in terms of the variable
$u=X/\rho=-\cos\theta$. Here
the atom lies on the line $y=0$, $x=-X$, and $\rho$ is the distance from the
edge of the plane and the atom. 
 $f>0$ is attractive, $f<0$ repulsive.
$\gamma$ goes from 0 to
1 by steps of 0.1, from bottom to top.
For $\gamma<1/4$ a repulsive regime always occurs when the atom is sufficiently
close to the plane of the conductor.
 }
\end{center}
\end{figure}

\begin{figure}
\begin{center}
\includegraphics[scale=1]{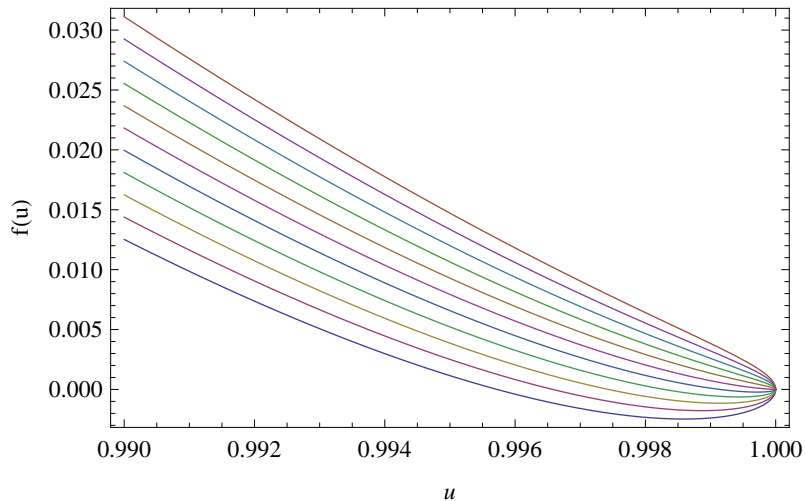}
\caption{\label{figsemi2}  The region close
to the plane, $1\ge u\ge 0.99$,
 with $\gamma$ near the critical value of 1/4.
Here from bottom to top are shown the results for values of $\gamma$ from
0.245 to 0.255 by steps of 0.001.}

\end{center}
\end{figure}

The energy for the completely anisotropic atom is
\be
V_{zz}=\frac13V(u)+\frac{u^4}2(1-3u^2).
\ee
If we consider instead a cylindrically symmetric polarizable
atom in which
\be
\bmalpha=\alpha_{zz}\mathbf{\hat z\hat z}+\gamma\alpha_{zz}(\mathbf{\hat x
\hat x+\hat y\hat y})=\alpha_{zz}(1-\gamma)\mathbf{\hat z\hat z}
+\gamma\alpha_{zz}{\bf1},\label{gammapol}
\ee
where $\gamma$ is the ratio of the transverse polarizability to the
longitudinal polarizability of the atom, the effective potential is
\be
(1-\gamma)V_{zz}+\gamma V,\label{effpot}
\ee
and the $z$-component of the force on the atom is
\be
F^\gamma_z=-\frac{\alpha_{zz}(0)}{32\pi}\frac1{X^5}u^2\sqrt{1-u^2}
\frac{d}{du}
\bigg[\frac12u^4(1-\gamma)(1-3u^2)+\frac13(1+2\gamma)V(u)\bigg],\label{fz}
\ee
where $V$ is given by (\ref{vee}).  Note 
that the energy, or the quantity in square brackets in (\ref{fz}), 
only vanishes at $u=1$ (the
plane of the conductor) when $\gamma=0$. Thus, the argument given in
Levin et al.\cite{levin} applies only for the completely anisotropic case.
Figure \ref{figsemi1} shows the dependence of $F_z$ on the polar angle.
Figure \ref{figsemi2} gives a finer resolution plot.
The critical value of $\gamma_c=1/4$ marks the boundary between the
regime where no repulsion occurs, and where repulsion occurs close to
the plane of the conductor.
It is interesting to observe that the same critical value of $\gamma$
occurs for the nonretarded (electrostatic) regime of a circular aperture, 
as follows from a simple computation based on the result of 
Eberlein and Zietal.\cite{zietal1,zietal2}
\bea
U&=&-\frac1{16\pi^2}\int_{-\infty}^\infty d\zeta\,\alpha_{zz}(\zeta)
\frac1{Z^3}\bigg\{(1+\gamma)\left(\frac\pi2+\arctan\frac{Z^2-a^2}{2aZ}\right)\nn
\\
&&\quad\mbox{}+\frac{2aZ}{(Z^2+a^2)^3}\left[(1+\gamma)(Z^4-a^4)
-\frac83(1-\gamma)a^2Z^2\right]\bigg\}.
\eea
It is easy to see that this has a minimum for $z>0$, and hence there is a repulsive
force close to the aperture, provided $\gamma<\gamma_c=1/4$.

\subsubsection{Repulsion by a wedge}
\begin{figure}
\begin{center}
\includegraphics[scale=1]{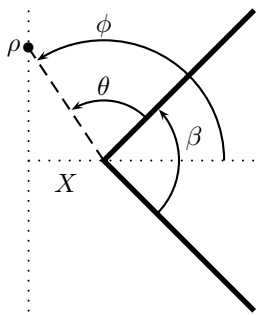}
\caption{\label{figw} Polarizable atom outside a perfectly conducting wedge
of opening angle $\beta$.}
\end{center}
\end{figure}

\begin{figure}
\begin{center}
\includegraphics[scale=.5]{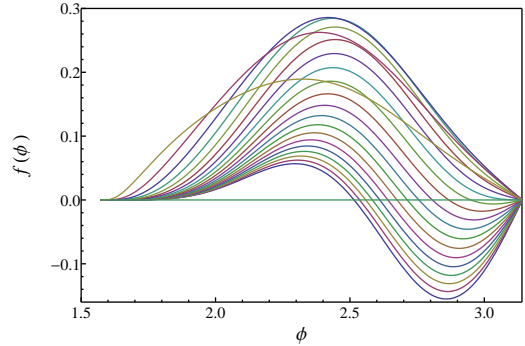}
\caption{\label{figww} $F_z$ for a completely anisotropic atom moving on
a line perpendicular to the wedge.  The different curves are for different
values of the interior angle $\beta=n\pi/20$, $n=0$ to 20, from bottom up.}
\end{center}
\end{figure}

It is very easy to generalize the above result for a wedge, $p>1/2$.  That is,
we want to consider a strongly anisotropic atom, with only $\alpha_{zz}$
significant, to the left of a wedge of interior angle
\be
\beta=2\pi-\Omega,
\ee
as shown in Fig.~\ref{figw}.
We want the $z$ axis to be perpendicular to the symmetry axis of the wedge
so the relation between the polar angle of the atom and the
angle to the symmetry line
is
$\phi=\theta+\beta/2$,
where, as before, $\theta$ is the angle relative to the top surface of the
wedge.  The CP energy
 is changed only by the replacement $\cos2\theta$ by
$\cos2\phi$, with no change in $\sin p\theta$.
How does repulsion depend on the wedge angle $\beta$?
Write for an atom on the line $x=-X$
\be
U^{zz}_{\rm CP}=-\frac{\alpha_{zz}(0)}{8\pi X^4}V(\phi),
\ee
where
\be
V(\phi)=\cos^4\phi\bigg[\frac{p^4}{\sin^4\frac\pi2\frac{\phi-\beta/2}{\pi
-\beta/2}}-\frac23\frac{p^2(p^2-1)}{\sin^2\frac\pi2\frac{\phi-\beta/2}{\pi
-\beta/2}}+\frac1{45}(p^2-1)(p^2+11)\cos2\phi\bigg].
\ee
At the point of closest approach,
\be
V(\pi)=\frac1{45}(4p^2-1)(4p^2+11),
\ee
so the potential vanishes at that point only for the half-plane case,
$p=1/2$. 
The force in the $z$ direction is
\be
F_z=-\frac{\alpha_{zz}}{8\pi}\frac1{X^5}\cos^2\phi\frac{\partial V(\phi)}{\partial\phi}.
\ee
Figure \ref{figww} shows the force as a function of $\phi$ for fixed $X$.
It will be seen that the force has a repulsive region for angles close
enough to the apex of the wedge, provided that the wedge angle is not too
large.  The critical wedge angle is actually
rather large, $\beta_c=1.87795$, or about
108$^\circ$.  For larger angles, the $z$-component of the force exhibits only
attraction.  

\subsection{CP repulsion by cylinder not sphere}
\begin{figure}
\begin{center}
\includegraphics{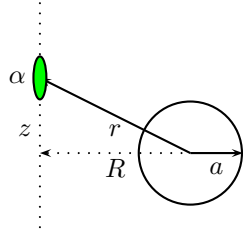}
\caption{\label{fig-cyl-atom} Interaction between an anisotropically polarizable
atom and a conducting cylinder of radius $a$.  The force on the atom along a line
which does not intersect the cylinder is considered.  If the atom is only
polarizable in that direction, and the line lies sufficiently far from the
cylinder, the force component along the line changes sign near the point of
closest approach.}
\end{center}
\end{figure}

\begin{figure}
\begin{center}
\includegraphics[scale=.5]{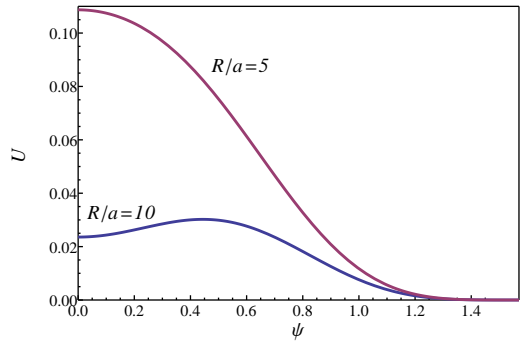}
\end{center}
\caption{\label{figcylatom} CP energy between a completely anisotropic atom
and a cylinder.  The motion of the atom, and its polarizability,
are along a line perpendicular to the cylinder, and not intersecting
with it.
Here $\psi$ is the angle above the radial line perpendicular to the line
 of motion of the atom.  That is, according to Fig.~\ref{fig-cyl-atom}, 
$\sin\psi=z/r$.}
\end{figure}

Finally, we turn to the Casimir-Polder interaction between an
anisotropic atom and an infinite cylinder, for the force on
the atom along a line perpendicular to, and not intersecting,
the cylinder, which is also polarizable only along the same
direction. The situation is illustrated inf Fig.~\ref{fig-cyl-atom}.
 The details will appear elsewhere.\cite{rep2}
Figure \ref{figcylatom} shows a plot of the CP  energy, the upper curve
being for the distance of closest approach to the cylinder axis
 $R$ being 5 times the cylinder
radius $a$, the lower curve for the distance of closest approach 10 times the
radius. Repulsion is clearly observed when $R/a=10$, but not for $R/a=5$.  In contrast,
for a conducting sphere, since at large distances it looks like an
isotropic polarizable
atom (with both electric and magnetic polarizabilities), no repulsion
on a completely anisotropic atom occurs.

\section{Conclusions}
\begin{itemlist}
 \item Casimir self-energies often exhibit repulsion, but
general systematics are not yet completely worked out.
\item Repulsion occurs between electric and magnetic conductors,
or materials or metamaterials that mimic this behavior over
a wide frequency range.  This is extraordinarily difficult to
achieve in practice.
\item Intervening intermediate ``density'' materials can mimic repulsion.
\item But true repulsion can be exhibited in Casimir-Polder
 situations with suitable
anisotropies.
\item New examples of Casimir and Casimir-Polder repulsion are still
being discovered.
\end{itemlist}

\section*{Acknowledgments}

We thank the US National Science Foundation, the US Department of
Energy, and the European Science Foundation for partial support of
this research.


\end{document}